\documentclass[12pt]{iopart}

\def\beqra{\begin{eqnarray}}
\def\eeqra{\end{eqnarray}}
\def\beq{\begin{equation}}
\def\eeq{\end{equation}}

\def\vp{\varphi}

\def\gmn{g_{\mu\nu}}
\def\tgmn{ \tilde{g}_{\mu\nu}}

\begin{document}
\title[Local transformations of units in Scalar-Tensor Cosmology]
{Local transformations of units in Scalar-Tensor Cosmology}

\author{R Catena$^1$, M Pietroni$^2$,L Scarabello$^3$}

\address{$^1$ Deutsches Elektronen-Syncrotron DESY, 22603 Hamburg, Germany}
\address{$^2$ INFN, Sezione di Padova, via Marzolo 8, I-35131, Padova, Italy}
\address{$^3$ Dipartimento di Fisica Universit\`a di Padova and INFN, Sezione di Padova, via Marzolo 8, I-35131, Padova, Italy}

\begin{abstract}
The physical equivalence of Einstein and Jordan frame in Scalar Tensor theories has been explained
by Dicke in 1962: they are related by a local transformation of units. We discuss this point
in a cosmological framework. Our main result is the construction of a formalism in which all the
physical observables are frame-invariant.  
The application of this approach to CMB codes is at present under analysis.
\end{abstract}


\section{Introduction}

Scalar-Tensor theories (ST) can be considered as the simplest extension of General Relativity (GR) that preserves
a universal metric coupling between gravity and matter \cite{bd, dam, GRattr}. In such theories the gravitational interaction
is described in terms of both a metric tensor and a scalar field. This feature, together with the relevance of scalar fields in Cosmology,
makes ST theories cosmologically interesting \cite{Bartolo, Catena, Esposito, Schimd, Perrotta, Matarrese}. 

ST theories can be formulated in different frames. 
In the so-called `Jordan frame', the Einstein-Hilbert action of GR is modified by the introduction 
of a scalar field $\vp$ with a non-canonical kinetic term and a potential. 
This field replaces the Planck mass, which becomes a dynamical quantity. 
On the other hand, the matter part of the action is just the standard one.

By Weyl-rescaling the Jordan metric $\tgmn$ as in the following 

\beq \tgmn = e^{-2 b(\vp)} \gmn\,,
\label{trans}
\eeq

one can express the ST action in the so called `Einstein Frame'. 
In the new variable $\gmn$, the gravitational action is just the Einstein-Hilbert one plus 
a scalar field with canonically normalized kinetic terms and an effective potential. 
On the other hand, in the matter action the scalar field appears, through the rescaling 
factor multiplying the metric tensor everywhere. As a consequence, 
the matter energy-momentum tensor is not covariantly conserved, 
and particle physics parameters, like masses and dimensionful coupling constants are space-time dependent.

The physical equivalence between the two frames was clearly discussed by Dicke in 1962 \cite{Dicke}.
He showed that the local Weyl rescaling of eq. (\ref{trans}) amounts to a local transformation of units;
as a result the physical equivalence between different frames is trivial. 
We develop here a formalism that makes transparent such an equivalence:
in our language all the observable quantities are manifestly frame-invariant 
(invariant under local transformations of units).  
Sections 1,2,3 and 4 are devoted to the construction of a frame-invariant action for ST theories.
In sections 5 and 6 we discuss such a formalism in a cosmological framework. 

\section{Local transformations of units}

We start summarizing Dicke's argument. Let's consider the separation $\bar{AB}$ between two spatial points A and B.
The measure of $\bar{AB}$ in the units $\mathbf{u}$ is given by $l\equiv \bar{AB}/\mathbf{u}$.
Under a transformation of units 

\beq
\mathbf{u} \to \lambda^{-1/2} \mathbf{u}
\label{u}
\eeq

the tranformation properties of $l$ are given by $l \to \lambda^{1/2} l$. 
The same argumet we just applied to a spatial interval could have been applied to 
a time interval. Therefore a space-time interval $ds = \sqrt{\gmn dx^{\mu}dx^{\nu}}$ transforms under a trasformation of units (\ref{u})
in the following way $ds \to \lambda^{1/2} ds$. We should now remeber that by definition the coordinates $x^{\rho}$ are invariant
under the transformation (\ref{u}). As a result, the metric tensor tranforms under the action of (\ref{u}) like

\beq
\gmn \to \lambda \gmn \,.
\eeq
 
Comparing now the last expression with eq. (\ref{trans}), we can conclude
that the local Weyl rescaling (\ref{trans}) has the meaning of a local transformation of units with $\lambda(x^{\rho}) = e^{-2b(\vp(x^{\rho}))}$. 
With similar arguments one could derive the transformation properties of all the important quantities. 
For example one has       

\beqra
&& l \to e^{-b(\vp)} \, l  \,,\;\;\; m \to e^{b(\vp)} \, m\,,\;\;\; \phi \to e^{b(\vp)} \, \phi \,,\;\;\;  \psi \to e^{3/2} \, \psi\,,\nonumber\\
&& A_\mu \to A_\mu\,,\,\;\;\;  P_\mu \to P_\mu\,,\,\;\;\; \Gamma \to e^{b(\vp)} \, \Gamma\,,\,\;\;\;  \dots
\label{ft}
\eeqra

for a length, a mass, a scalar field, a spinor, a vector, the canonic momenta and a rate respectively.
It should be clear now that in the language of Dicke to choose a frame amounts to a local choice of units.
 
\section{Frame-invariant variables}

Knowing the transformation properties of the important quantities, one can define a set of frame-invariant 
variables through which to construct a frame-invariant action for ST theories. Let's start introducing
a reference length $l_{R}$. Its transformation properties under local transformations of units are specified
in eq. (\ref{ft}). With this object we can construct, for example, the following frame-invariant quantities

\beqra
&& \bar{m} \equiv l_{R} \, m\,,\;\;\; \bar{\phi} \equiv l_{R} \, \phi \,,\;\;\;  \bar{\psi} \equiv l^{3/2}_{R} \, \psi\,,\nonumber\\
&& \bar{\Gamma} \equiv l_{R} \, \Gamma\,,\,\;\;\; h_{\mu \nu} \equiv l^{-2}_{R} \gmn \,,\,\;\;\; \dots
\label{fi}
\eeqra

We will discuss in more details the properties of the metric $h_{\mu \nu}$ when we will embed our discussion in
a cosmological framework. We want to stress here that $l_{R}$ is not a new scale of the theory and it's not a dynamical
field. From a mathematical point of view, 
it's just a quantity transforming like a length under a local transformation of units that we use for
the definitions (\ref{fi}). 
Physically, its meaning becomes clear if one observes that only ratio between physical scales
are really accessible to the experiments: $l_{R}$, therefore, represents a convenient reference magnitude
appearing in such a ratio at the moment of performing a measurement.  
As a result, the choice of its value is dictated by practical convenience. 
For instance, in astrophysics $l_{R}$ can be a reference atomic wavelength, in particle physics 
the inverse of some particle mass, or, in gravitational theories, the Planck length.
ST theories are characterized by the possibility to use space-time dependent reference lenghts. 
As we will see in the next section, in our language to choose a frame amounts to a choice of the function $l_{R}(x^{\rho})$.

\section{Frame-invariant action}

Scalar-tensor theories can be defined in terms of frame-independent quantities by the action

\beq
S=S_G[h_{\mu\nu},\,\varphi] + S_M[h_{\mu\nu} e^{-2 b(\vp)},\,\bar{\phi}\,,\bar{\psi}\,,\ldots]\,,
\label{action}
\eeq

where the frame-independent fields $\bar{\phi}\,,\bar{\psi}\,,\ldots$, 
appearing in the matter action $S_M$ are given by the combinations in eq.~(\ref{fi}). 
The gravity action is given by

\beq
S_G= \kappa \int d^4 x \,\sqrt{-h}\left[ R(h) - 2 \,h^{\mu\nu} \partial_\mu\vp \,\partial_\nu\vp-4 U(\vp)\right]\,.
\label{actiongrav}
\eeq

In our language, choosing a frame corresponds to fixing the function $l_{R}(x^{\rho})$ appropriately. 
The first possible choice is to take a constant $l_{R}(x^{\rho}) = l_{Pl}$, which corresponds to the Einstein frame. 
The gravity action takes  the usual Einstein-Hilbert form

\beq
S_G= \kappa\, l_{Pl}^{-2} \int d^4 x \,\sqrt{-g}\left[ R(g) - 2 \,g^{\mu\nu} \partial_\mu\vp \,\partial_\nu\vp-4 V(\vp)\right]\,,
\eeq

with $V=l_{Pl}^{-2} \,U$. The combination in front of the integral fixes the Einstein-frame Planck mass, 
$\kappa\, l_{Pl}^{-2} = M_\ast^2/2 =(16 \pi G_\ast)^{-2}$. The matter action is obtained from the one of quantum field theory by substituting the 
Minkowsky metric $\eta_{\mu\nu}$ with $g_{\mu\nu} e^{-2b(\vp)}$. Since in this frame the matter energy-momentum tensor 
is not conserved, particle physics quantities, like masses and wavelengths are not constant. \\
The other choice corresponds to the Jordan frame, which is obtained if one choses 
$l_{R}(x^{\rho}) = \tilde{l}_{P}(x^{\rho}) = l_{Pl}\, e^{-b(\vp(x^{\rho}))}$, 
where $l_{Pl}$ is the previously defined Planck length in the Einstein frame. With this choice the matter action 
takes the standard form of quantum field theory (with $\eta_{\mu\nu}\rightarrow g_{\mu\nu}$), whereas the gravity action is

\beq
S_G = \frac{M_\ast^2}{2} \int d^4 x \,\sqrt{-\tilde{g}} \,e^{2b(\vp)} \left[ R(\tilde{g}) 
- 2 \,\tilde{g}^{\mu\nu} \partial_\mu\vp \,\partial_\nu\vp \,(1-3 \,\alpha^2)-4 \tilde{V}(\vp)\right]\,,
\eeq

where $\tilde{V}=\tilde{l}_{P}^{-2}U$, $\tgmn$ is defined in (\ref{trans}) and $\alpha\equiv \frac{d b}{d\vp}$.
Notice that, in this frame, the r\^{o}le of the Planck mass is played by the space-time dependent quantity 
$M_\ast e^{b(\vp)}$. Since $b(\vp)$ disappears from the matter action, the energy-momentum tensor is now covariantly conserved.

\section{Background cosmology}

Assuming a FRW structure for the metric $\gmn$ in eq. (\ref{fi}) we can consistently write the
frame-invariant metric $h_{\mu \nu}$ in the following way

\beq 
dh^2= -a^2(\tau) \langle l \rangle ^{-2}_{R} (\tau)  (d\tau^2- \delta_{ij}dx^i dx^j)\,,
\label{m}
\eeq

where $dh^2 = h_{\mu \nu} dx^{\mu} dx^{\nu}$ and we took in to account that in a generic frame
$l_{R} (x^{\rho}) =  \langle l \rangle_{R}(\tau) + \delta l_{R}(x^{\rho})$ can eventually be space-time dependent.
The quantity $\langle l \rangle_{R}(\tau)$ represents a spatial average on a time-slice for an observer
in the CMB rest-frame.\\
The metric (\ref{m}) implies the following redshift-scale factor relation

\beq
1+z(\tau)  = \frac{a(\tau_0)}{a(\tau)}\frac{\bar{l}_{at}(\tau)}{\bar{l}_{at}(\tau_0)} \;.
\label{red}
\eeq

Here we wrote $l_{R}=l_{at}$ to underline that in redshift measurement an atomic wavelength is tipically used as reference length. 
The standard relation between the redshift and the scale factor, {\it i.e.} $1+z = a(\tau_0)/a(\tau)$ 
is recovered only in that frame in which the reference wavelength $l_{at}$ is constant in time and space. 
In ST theories this is the case of the Jordan frame.\\
Background fluids motion is described in the frame invariant phase space $(x^{i}, P_{j})$.
On this phase space one can define a frame-invariant distribution function $F(x^{i}, P_{j}, \tau)$ and consequently a frame-invariant
energy-momentum tensor

\beq 
\bar{T}_{\mu\nu} = l_{R}^4g_s \int \frac{d^3P}{(2\pi)^3} (-g)^{-1/2}\,\frac{P_\mu P_\nu}{P^0} F(x^{i},P_{,}, \tau) \,,
\eeq

where $d^3P = dP_1 dP_2 dP_3$ and $g_s$ counts the spin degrees of freedom.

From the last expression we can read the frame-invariant energy density $\bar{\rho} = -\bar{T}^{0}_{0}$ and pressure 
$\bar{p} \, \delta^{i}_{j} = -\bar{T}^{i}_{j}$. By variation of eq. (\ref{action}) 
is now possible to derive the frame-invariant equations of motion in a consistent frame-invariant FRW background. 
The results are given in \cite{EJ}.

\section{Cosmological perturbations}

We extend here the formalism we developed so far to first order in perturbation theory.
Let's start including first order perturbations of the metric (\ref{m})

\beqra
&&
dh^2 =  a^2(\tau)/\langle l \rangle_R^2 \left[-\left(1+2\,\Psi - 2\,\frac{\delta l_R}{l_R}\right)d\tau^2 \right.\nonumber\\
&& \qquad\qquad \qquad\quad\qquad\qquad\left.+ \left(1-2\,\Phi- 2\,\frac{\delta l_R}{l_R}\right)\delta_{ij}dx^i dx^j\right]\,.
\label{mi1}
\eeqra

From eq. (\ref{mi1}) we can read the definitions of frame-invariant scale factor and potentials

\beq
\bar{a}\equiv a/\bar{l}_R\,,\;\;\;\;\;\bar{\Psi}\equiv \Psi -\frac{\delta l_R}{l_R}\,,\;\;\;\;\;\bar{\Phi}\equiv\Phi +\frac{\delta l_R}{l_R}\,.
\label{fisp}
\eeq

The first order equations of motions obtained by eqs. (\ref{action}) and (\ref{mi1}) are given in \cite{EJ}. \\
In the next section we will apply the formalism we developed so far to the Boltzmann equation.
This can be of interest in many fields of Cosmology.

\section{The Boltzmann equation}

The evolution of the phase space density of a particle $\psi$, $F_\psi(x_\psi^i,P^\psi_j,\tau)$ is given by
the Boltzmann equation

\beq
\frac{\partial F^\psi}{\partial \tau} +\frac{d x_\psi^i}{d \tau} 
\frac{\partial F^\psi}{\partial x_\psi^i} +\frac{d P^\psi_j}{d \tau} 
\frac{\partial F^\psi}{\partial P^\psi_j} = \left[\frac{d F^\psi}{d \tau} \right]_C\,.
\label{bol}
\eeq

The frame-invariance of the LHS is triavially checked. Working with our variables, 
it's an immediate consequence of the manifest frame-invariance of the geodesic equation \cite{EJ}

\beq
\frac{dP^{\psi}_j}{d\tau} - P^{\psi}_0 \,\partial_j(\bar{\Psi}+\log\,\bar{m}) = 0\,.
\label{geod}
\eeq

The collisional term for a generic process $\psi+a+b+\cdots\leftrightarrow i+j+\cdots$ reads

\beqra
&& \left[\frac{d F_\psi}{d \tau} \right]_C(x_\psi^i,P^\psi_j,\tau)\nonumber =\frac{1}{2P^\psi_0}
 \int d \Pi^a d \Pi^b\cdots d \Pi^i d \Pi^j\cdots \nonumber\\
 &&\times (2\pi)^4 \delta^4(P^\psi+P^a+P^b \cdots -P^i-P^j\cdots)\nonumber\\
&& \times\left[|{\cal M}|_{\psi+a+b+\cdots\rightarrow i+j+\cdots}^2   F_\psi F_a F_b\cdots (1\pm F_i) (1\pm F_j)\cdots \right.\nonumber\\
&& \left.- |{\cal M}|_{i+j+\cdots\rightarrow \psi+a+b+\cdots }^2 F_i F_j\cdots (1\pm F_\psi)(1\pm F_a)(1\pm F_b) \cdots \right]\,,
\label{collision}
\eeqra

where the $''+''$ applies to bosons and the $''-''$ to fermions, and $d \Pi$ is the frame-invariant quantity

\beqra
&& d \Pi \equiv l_R^2 \,\frac{d^4P}{(2\pi)^3}\,(-g)^{-1/2} \, \delta(P^2+m^2) \Theta(P^0)\nonumber\\
&&\quad= l_R^2\, \frac{d^3P}{(2\pi)^3}\frac{(-g)^{-1/2}}{2P^0} \,,
\eeqra

with $d^4P = dP_0 \, d^3P$.
The delta-function in eq.~(\ref{collision}) depends on momenta with low indices.

\section{Conclusions}

Following Dicke's argument \cite{Dicke}, we discussed here the relation between different frames
in ST theories in terms of local trasformations of units. In this language the physical equivalence
of Einstein and Jordan frame is manifest and all the physical observables can be written in terms of
frame-invariant quantities. We applied such a formalism to the first order perturbed Boltzmann equation.
Its frame-invariant expression (\ref{bol}) can be of interest in many analytical and numerical computations.

\section*{Acknowledgments} 

R. Catena acknowledges a Research Grant funded by the VIPAC Institute.

\end{document}